\documentclass[a4paper,12pt]{article}

\usepackage{graphicx}
\usepackage{amsmath}
\usepackage{amssymb}
\usepackage{hyperref}
\usepackage{tabularx}
\newcolumntype{M}{>{$}c<{$}}
\usepackage[matrix,arrow,curve]{xy}

\numberwithin{equation}{section} \numberwithin{figure}{section}
\numberwithin{table}{section}

\def\papertitlepage{\baselineskip 3.5ex\thispagestyle{empty}}
\def\Title#1{\baselineskip 1cm \vspace{1.5cm}%
  \begin{center}{\Large\bf #1}\end{center}\vspace{0.5cm}}
\def\Authors#1{\begin{center}\renewcommand{\thefootnote}{\fnsymbol{footnote}}{\it #1}\end{center}}
\def\Abstract{\vspace{1.0cm}%
  \begin{center}{\large\bf Abstract}\end{center}}

\renewenvironment{thebibliography}{\pagebreak[3]\par\vspace{0.6em}
\begin{flushleft}{\large \bf References}\end{flushleft}
\vspace{-1.0em}

\begin{enumerate}\if@twocolumn\baselineskip=0.6em\itemsep -0.2em
\else\itemsep -0.2em\fi\labelsep 0.1em}{\end{enumerate} }

\DeclareMathDelimiter{\lcolon}{\mathopen}{operators}{"3A}{largesymbols}{"3A}
\DeclareMathDelimiter{\rcolon}{\mathclose}{operators}{"3A}{largesymbols}{"3A}

\def\+{\!\!+\!\!}

\def\dynkin(#1){(#1)}

\def\bra<#1|{\langle#1|}
\def\ket|#1>{|#1\rangle}
\def\braket<#1|#2>{\langle#1|#2\rangle}
\def\llangle{\langle\!\langle}
\def\rrangle{\rangle\!\rangle}
\def\bbra<#1|{\llangle#1|}
\def\kket|#1>{|#1\rrangle}
\def\bbraket<#1|#2>{\llangle#1|#2\rrangle}

\begin{document}
{\papertitlepage \vspace*{0cm} {\hfill
\begin{minipage}{4.2cm}
IF-USP 2012\par\noindent May, 2012
\end{minipage}}
\Title{Multibrane solutions in cubic superstring field theory}
\Authors{{\sc E.~Aldo~Arroyo${}$\footnote{\tt
aldohep@fma.if.usp.br}}
\\
Instituto de F\'{i}sica, Universidade de S\~{a}o Paulo \\[-2ex]
C.P. 66.318 CEP 05315-970, S\~{a}o Paulo, SP, Brasil ${}$ }
} 

\vskip-\baselineskip
{\baselineskip .5cm \Abstract Using the elements of the so-called
$KBc\gamma$ subalgebra, we study a class of analytic solutions
depending on a single function $F(K)$ in the modified cubic
superstring field theory. We compute the energy associated to
these solutions and show that the result can be expressed in terms
of a contour integral. For a particular choice of the function
$F(K)$, we show that the energy is given by integer multiples of a
single D-brane tension.

 }
\newpage
\setcounter{footnote}{0}
\tableofcontents

\section{Introduction}
After the discovery of the first analytic solution for tachyon
condensation in open bosonic string field theory
\cite{Schnabl:2005gv}, there has been a remarkable amount of work
regarding the analytic set up of the theory. One line of research
is concerned with the algebraic structure of the string field
algebra based on Witten's associative star product
\cite{Witten:1985cc,Okawa:2006vm,Erler:2006hw,Erler:2006ww,Zeze:2011zz,Schnabl:2010tb}.
Many authors have used the so-called $KBc$ subalgebra to
understand systematically the construction of analytic solutions
and perform computations by means of purely algebraic
manipulations
\cite{Erler:2009uj,Zeze:2010jv,Zeze:2010sr,Arroyo:2010sy,AldoArroyo:2009hf,Erler:2012qr}.

Consequently, using the elements of the $KBc$ subalgebra, it was
possible to rewrite many gauge equivalent tachyon vacuum solutions
\cite{Erler:2006ww}. These solutions have been successfully used
to prove Sen's conjecture \cite{Sen:1999mh,Sen:1999xm}. Moreover,
the possibility of constructing another set of solutions besides
the well known tachyon vacuum solution was studied
\cite{Ellwood:2009zf,Erler:2012qn,Bonora:2011ns,Bonora:2011ru,Erler:2011tc,Takahashi:2011wk,Hata:2011ke}.
In recent work \cite{Murata:2011ex,Murata:2011ep}, the possibility
of describing multibrane configurations by employing a class of
analytic solutions of the string field equation of motion in open
bosonic string field theory was discussed.

In the context of the modified cubic string field theory
\cite{Arefeva:1989cp}, the analytic construction of the tachyon
vacuum solution was analyzed first by Erler \cite{Erler:2007xt}.
Then further discussions were given in a set of papers
\cite{Aref'eva:2008ad,Gorbachev:2010zz,Aref'eva:2009sj,Arefeva:2010yd,Kroyter:2009bg},
where the so-called $KBc\gamma$ subalgebra was introduced
\cite{Arroyo:2010fq,Erler:2010pr}. Many gauge equivalent tachyon
vacuum solutions were discovered and the computation of the energy
associated to these solutions was performed analytically and
numerically given results in agreement with Sen's conjecture.
However, the description of multibrane configurations by means of
some general set of solutions constructed out of elements in the
$KBc\gamma$ subalgebra was not considered.

In this paper, we explore the possibility of describing multibrane
configurations by constructing a class of analytic solutions of
the string field equation of motion in the modified cubic
superstring field theory. These solutions will be expressed in
terms of elements in the $KBc\gamma$ subalgebra. As discussed in
reference \cite{Arroyo:2010fq}, there is a well established
prescription to find solutions which follows two steps: (i) find a
naive identity based solution of the string field equation of
motion, and (ii) perform a gauge transformation on the identity
based solution such that the resulting string field $\Psi$
unambiguously reproduces a finite value for the energy computed
from the cubic string field action
\begin{eqnarray}
\label{poten1intro} U(\Psi)=\frac{1}{2} \langle \Psi Q \Psi
\rangle+\frac{1}{3} \langle \Psi \Psi \Psi \rangle  \, ,
\end{eqnarray}
where $Q$ is the BRST operator of the open Neveu-Schwarz
superstring theory. In the correlator $ \langle \cdot \cdot \cdot
\rangle $ we must insert the operator $Y_{-2}$ at the open string
midpoint. The operator $Y_{-2}$ can be written as the product of
two inverse picture changing operators $Y_{-2}=Y(i)Y(-i)$, where
$Y(z)=-\partial \xi e^{-2 \phi} c(z)$. The string field $\Psi$
which has ghost number 1 and picture number 0 belongs to the small
Hilbert space of the first-quantized matter+ghost open
Neveu-Schwarz superstring theory.

In the case of the modified cubic superstring field theory, in
addition to the basic string field elements $K$, $B$ and $c$, we
need to include the super-reparametrization ghost field $\gamma$
\cite{Arroyo:2010sy,Erler:2007xt,Gorbachev:2010zz,Arroyo:2010fq}.
These basic string fields satisfy a set of algebraic relations,
and with the help of these relations we can construct the
following identity based solution
\begin{eqnarray}
\label{02eq1intro} \Psi_I=(c+B \gamma^2)(1-K),
\end{eqnarray}
which formally satisfies the string field equation of motion $Q
\Psi_I + \Psi_I \Psi_I =0$ of the cubic theory. This solution was
shown to be gauge equivalent \cite{Arroyo:2010fq} to the solution
found by Gorbachev \cite{Gorbachev:2010zz}.

A class of analytic solutions of the string field equation of
motion in the modified cubic superstring field theory can be
derived by performing a rather general gauge transformation on the
identity based solution
\begin{eqnarray}
\label{solpsi1intro} \Psi=
\mathcal{U}_F(Q+\Psi_I)\mathcal{U}^{-1}_F,
\end{eqnarray}
where the element of the gauge transformation can be explicitly
constructed in terms of the basic string fields $K$, $B$ and $c$
\begin{eqnarray}
\label{gaugeU1intro} \mathcal{U}_F =
F\Big(1+cB\frac{K-1+F^2}{1-F^2}\Big), \;\;\;\;\;
\mathcal{U}^{-1}_F =\Big(1-cB\frac{K-1+F^2}{K}\Big)\frac{1}{F}
\end{eqnarray}
with $F=F(K)$ being an arbitrary function of $K$.

Carrying out a suitable algebraic manipulations in the $KBc\gamma$
subalgebra, from the gauge transformation (\ref{gaugeU1intro}), we
obtain the following set of solutions that depend on the single
function $F(K)$
\begin{eqnarray}
\label{solpsi2intro} \Psi= Fc\frac{KB}{1-F^2}cF +F B\gamma^2 F.
\end{eqnarray}
This solution was analyzed in references
\cite{Erler:2007xt,Gorbachev:2010zz} for the particular cases:
$F^2=e^{-K}$ and $F^2=1/(1+K)$, where it was shown that the
solutions describe the tachyon vacuum solution. Discussions
related to the gauge equivalence of these solutions were given in
reference \cite{Arefeva:2010yd}. Nevertheless, there had been no
evaluation of the energy for a class of analytic solutions of the
form (\ref{solpsi2intro}) for a generic function $F(K)$.

In order to compute the energy for the solution given by
(\ref{solpsi2intro}), it should be convenient to define the
function $G(K)=1-F^2(K)$. Under certain holomorphicity conditions
satisfied by the function $G(K)$, we will show that the expression
for the energy can be written in terms of a contour integral
\begin{eqnarray}
\label{kine1x6intro} U(\Psi) = - \frac{1}{2 \pi^2} \oint
\frac{dz}{2 \pi i}  \frac{ G'(z)}{G(z)}.
\end{eqnarray}
To compute this integral (\ref{kine1x6intro}), we need to consider
a contour encircling the origin in the counterclockwise direction.

A function $G(K)$ which satisfies the holomorphicity conditions
analyzed in this paper is given by
\begin{eqnarray}
\label{GGintro} G(K)=\Big( \frac{K}{1+K} \Big)^n,
\end{eqnarray}
where $n$ is a non-negative integer, this is because it will count
the number of D-branes.

Since the contour integral (\ref{kine1x6intro}) is performed
around a closed curve encircling the origin, to compute the
integral we need to write the Laurent series of the integrand
around $z=0$ and pick up the coefficient in front of the term
$1/z$. For the function $G(K)$ defined by equation
(\ref{GGintro}), it turns out that
\begin{eqnarray}
\label{expandg}\frac{ G'(z)}{G(z)} = \frac{n}{z} + \sum_{m \neq
-1} b_m z^m,
\end{eqnarray}
and consequently the contour integral (\ref{kine1x6intro}) gives
the following result
\begin{eqnarray}
\label{multipotencilalx1yy1} U(\Psi) = - \frac{n}{2 \pi^2},
\end{eqnarray}
which is the expected result for a multibrane solution. Therefore,
in the context of the modified cubic superstring field theory, as
in the case of open bosonic string field theory, we should also
expect a solution which describes the so-called multibrane
configuration \cite{Murata:2011ex,Murata:2011ep}.

This paper is organized as follows. In section 2, we study a class
of analytic solutions of the string field equation of motion in
the modified cubic superstring field theory. By performing an
explicit gauge transformation, we show that these analytic
solutions depending on a single function $F(K)$ can be related to
an identity based solution. In section 3, by considering a generic
function $F(K)$, we evaluate the energy associated to the analytic
solution derived in the previous section. In section 4, for a
particular choice of the function $F(K)$, we show that the energy
of the solution is given by integer multiples of a single D-brane
tension. In section 5, a summary and further directions of
exploration are given.

\section{Derivation of the solution}
In this section, we are going to derive a rather general solution
to the string field equation of motion in the modified cubic
superstring field theory \cite{Arefeva:1989cp}. Using the
relations satisfied by the elements of the so-called $KBc\gamma$
subalgebra, the solution will be constructed by performing a gauge
transformation on an identity based solution.

Let us remember that, in the superstring case, in addition to the
basic string field elements $K$, $B$ and $c$, we need to include
the super-reparametrization ghost field $\gamma$
\cite{Arroyo:2010sy,Erler:2007xt,Gorbachev:2010zz,Arroyo:2010fq}.
These basic string fields satisfy the algebraic relations
\begin{align}
&\{B,c\}=1\, , \;\;\;\;\;\;\; [B,K]=0 \, , \;\;\;\;\;\;\;
B^2=c^2=0
\, , \nonumber\\
\label{02eq2} \partial c = [K&,c] \, , \;\;\;\;\;\;\;
\partial \gamma  = [K,\gamma] \, , \;\;\;\;\;\;\; [c,\gamma]=0 \, ,
\;\;\;\;\;\;\; [B,\gamma]=0 \, ,
\end{align}
and have the following BRST variations
\begin{eqnarray}
\label{02eq3} QK=0 \, , \;\;\;\;\;\; QB=K \, , \;\;\;\;\;\;
Qc=cKc-\gamma^2 \, , \;\;\;\;\;\; Q\gamma=c \partial \gamma
-\frac{1}{2} \gamma
\partial c \, .
\end{eqnarray}
Employing these basic string fields, we can construct the
following identity based solution
\begin{eqnarray}
\label{02eq1} \Psi_I=(c+B \gamma^2)(1-K),
\end{eqnarray}
which formally satisfies the string field equation of motion $Q
\Psi_I + \Psi_I \Psi_I =0$, where in this case $Q$ is the BRST
operator of the open Neveu-Schwarz superstring theory.

With the help of this algebraic construction, let us derive a
solution of the string field equation of motion by performing a
gauge transformation on the identity based solution
$\Psi_I=(c+B\gamma^2)(1-K)$
\begin{eqnarray}
\label{solpsi1} \Psi= \mathcal{U}_F(Q+\Psi_I)\mathcal{U}^{-1}_F,
\end{eqnarray}
where $\mathcal{U}_F$ is an element of the gauge transformation
given by
\begin{eqnarray}
\label{gaugeU1} \mathcal{U}_F =
F\Big(1+cB\frac{K-1+F^2}{1-F^2}\Big), \;\;\;\;\;
\mathcal{U}^{-1}_F =\Big(1-cB\frac{K-1+F^2}{K}\Big)\frac{1}{F}
\end{eqnarray}
with $F$ being a function of $K$.

Replacing (\ref{gaugeU1}) into (\ref{solpsi1}) and using the
identity based solution $\Psi_I=(c+B\gamma^2)(1-K)$, it is almost
easy to derive the following solution
\begin{eqnarray}
\label{solpsi2} \Psi= Fc\frac{KB}{1-F^2}cF +F B\gamma^2 F.
\end{eqnarray}
In the context of the modified cubic superstring field theory,
this solution was analyzed in references
\cite{Erler:2007xt,Gorbachev:2010zz} for the particular cases:
$F^2=e^{-K}$ and $F^2=1/(1+K)$, where it was shown that the
solutions describe the tachyon vacuum solution. Discussions
related to the gauge equivalence of these solutions were given in
reference \cite{Arefeva:2010yd}. Nevertheless, there had been no
evaluation of the energy for a class of analytic solutions of the
form (\ref{solpsi2}) for a generic function $F(K)$. In the next
section, we are essentially going to perform that calculation.

\section{Computation of the energy}

In this section, by considering a generic function $F(K)$, we are
going to evaluate the energy of the analytic solution derived in
the previous section. Let us mention that a similar computation
was performed by Murata and Schnabl in the context of open bosonic
string field theory \cite{Murata:2011ex,Murata:2011ep}. The energy
of a solution of the form (\ref{solpsi2}) can be computed from the
kinetic term. After some simplifications, we obtain
\begin{eqnarray}
\label{kine1} \langle \Psi Q\Psi \rangle = 2 \Big\langle
\frac{K}{G},(1-G),K,(1-G)\Big\rangle -2\Big\langle
\frac{K}{G},(1-G),\frac{K}{G},(1-G)\Big\rangle \\
- \Big\langle K,\frac{K}{G},(1-G),(1-G)\Big\rangle \nonumber +
\Big\langle \frac{K}{G},K,(1-G),(1-G)\Big\rangle,
\end{eqnarray}
where
\begin{eqnarray}
\label{G1} G= 1-F^2,
\end{eqnarray}
and to simplify the notation, we have defined
\begin{eqnarray}
\label{corre1} \Big\langle F_1,F_2,F_3,F_4\Big\rangle =
\langle\langle B F_1(K) c F_2(K) c
F_3(K)\gamma^2F_4(K)\rangle\rangle
\end{eqnarray}
for general $F_i(K)$. The insertion of notation $\langle\langle
\,\cdots \rangle\rangle$ stands for a standard correlator with the
difference that we must insert the operator $Y_{-2}$ at the open
string midpoint. The operator $Y_{-2}$ can be written as the
product of two inverse picture changing operators, $Y_{-2} = Y
(i)Y (-i)$, where $Y (z) = - \partial \xi e^{-2\phi}c(z)$.

Let us assume that all functions $F_i(K)$ can be written as a
continuous superposition of wedge states,
\begin{eqnarray}
\label{fi1} F_i(K)=\int_{0}^{\infty} d\alpha_i f_i(\alpha_i)
e^{-\alpha_i K}.
\end{eqnarray}
The validity of this assumption depends on some holomorphycity
conditions satisfied by the functions $F_i(K)$. These requirements
were analyzed in \cite{Schnabl:2010tb}. For a moment, let us
implicitly suppose that the functions  $F_i(K)$ satisfied the
aforementioned requirements.

Plugging the integral representation of the functions $F_i$'s
(\ref{fi1}) into (\ref{corre1}), we obtain the following quadruple
integral
\begin{align}
\label{corre1f2} \Big\langle F_1,F_2,F_3,F_4\Big\rangle =
\int_{0}^{\infty} d\alpha_1 d\alpha_2 d\alpha_3 d\alpha_4
f_1(\alpha_1) f_2(\alpha_2) f_3(\alpha_3)
f_4(\alpha_4)\langle\langle B e^{-\alpha_1 K}ce^{-\alpha_2
K}ce^{-\alpha_3 K}\gamma^2e^{-\alpha_4 K} \rangle\rangle
\end{align}
with the basic correlator $\langle\langle B e^{-\alpha_1
K}ce^{-\alpha_2 K}ce^{-\alpha_3 K}\gamma^2e^{-\alpha_4 K}
\rangle\rangle$ given by
\cite{Erler:2007xt,Gorbachev:2010zz,Arroyo:2010fq}
\begin{eqnarray}
\label{corref13} \langle\langle B e^{-\alpha_1 K}ce^{-\alpha_2
K}ce^{-\alpha_3 K}\gamma^2e^{-\alpha_4 K} \rangle\rangle =
\frac{s}{2 \pi^2} \alpha_2, \;\; \text{where} \;\; s=
\alpha_1+\alpha_2+\alpha_3+\alpha_4.
\end{eqnarray}

In what follows, we are going to use the $s$-$z$ trick introduced
in \cite{Murata:2011ex,Murata:2011ep}. Basically the trick
instructs us to insert the identity
\begin{eqnarray}
\label{ident1} 1=\int_{0}^{\infty} ds
\delta\Big(s-\sum_{i=1}^{4}\alpha_i\Big) =\int_{0}^{\infty}
ds\int_{-i\infty}^{+i\infty} \frac{dz}{2 \pi i} e^{sz}
e^{-z\sum_{i=1}^{4}\alpha_i},
\end{eqnarray}
into the quadruple integral (\ref{corre1f2}). This identity allows
us to treat the variable $s$ as independent of the other
integration variables $\alpha_i$. Using the correlator
(\ref{corref13}) and plugging the identity (\ref{ident1}) into
(\ref{corre1f2}), we get
\begin{align}
\label{corre1f2x1} \frac{1}{2 \pi^2}\int_{0}^{\infty} d\alpha_1
d\alpha_2 d\alpha_3 d\alpha_4 f_1(\alpha_1) \alpha_2 f_2(\alpha_2)
f_3(\alpha_3) f_4(\alpha_4) \int_{0}^{\infty}
ds\int_{-i\infty}^{+i\infty} \frac{dz}{2 \pi i} s \, e^{sz}
e^{-z\sum_{i=1}^{4}\alpha_i}.
\end{align}
Performing the integral over the variables $\alpha_i$ and
reexpressing the result in terms of the original functions
$F_i(z)$, we obtain
\begin{align}
\label{corre1f2x2}\Big\langle F_1,F_2,F_3,F_4\Big\rangle =
-\frac{1}{2 \pi^2} \int_{0}^{\infty} ds\int_{-i\infty}^{+i\infty}
\frac{dz}{2 \pi i} s \, e^{sz} F'_2(z)F_1(z)F_3(z)F_4(z).
\end{align}

With the help of this formula (\ref{corre1f2x2}), we are ready to
evaluate the kinetic energy $\langle \Psi Q\Psi \rangle$. For
instance, for the first term on the right-hand side of equation
(\ref{kine1}) the functions $F_i$'s are given by $F_1=K/G$,
$F_2=(1-G)$, $F_3=K$ and $F_4=(1-G)$, so that using
(\ref{corre1f2x2}), we arrive at the following result
\begin{eqnarray}
\label{extra1} \Big\langle
\frac{K}{G},(1-G),K,(1-G)\Big\rangle=-\frac{1}{2 \pi^2}
\int_{0}^{\infty} ds\int_{-i\infty}^{+i\infty} \frac{dz}{2 \pi i}
s \, e^{sz} z^2 \Big[ G'(z)-\frac{ G'(z)}{G(z)} \Big].
\end{eqnarray}

Performing a similar computation for the rest of terms on the
right-hand side of equation (\ref{kine1}), and adding up the
results, we derive an expression for the kinetic energy given by
\begin{eqnarray}
\label{kine1x2} \langle \Psi Q\Psi \rangle = -\frac{1}{2 \pi^2}
\int_{0}^{\infty} ds\int_{-i\infty}^{+i\infty} \frac{dz}{2 \pi i}
s \, e^{sz} z^2 \Big[ -\frac{6 G'(z)}{G(z)}+\frac{3
G'(z)}{G(z)^2}+3 G'(z) \Big].
\end{eqnarray}
Evaluating the integral over the variable $s$, which is well
defined for Re$(z)<0$, we obtain
\begin{eqnarray}
\label{kine1x3} \langle \Psi Q\Psi \rangle = -\frac{1}{2 \pi^2}
\lim_{\epsilon \rightarrow
0^{+}}\int_{-i\infty-\epsilon}^{+i\infty-\epsilon} \frac{dz}{2 \pi
i} \Big[ -\frac{6 G'(z)}{G(z)}+\frac{3 G'(z)}{G(z)^2}+3 G'(z)
\Big].
\end{eqnarray}

Let us suppose that the function $G$ can be expanded as
$G(z)=1+\sum_{n=1}^{\infty} a_n z^{-n}$, i.e., $G$ is holomorphic
at the point at infinity $z=\infty$ and has a limit $G(\infty)=1$.
Using this condition, we can make the integral along the imaginary
axis into a sufficiently large closed contour $C$ running in the
counterclockwise direction by adding a large non-contributing
half-circle in the left half plane Re$(z) < 0$. Therefore under
this assumption, the kinetic energy (\ref{kine1x3}) can be written
as
\begin{eqnarray}
\label{kine1x4} \langle \Psi Q\Psi \rangle = -\frac{1}{2 \pi^2}
\oint_{C} \frac{dz}{2 \pi i} \Big[ -\frac{6 G'(z)}{G(z)}+\frac{3
G'(z)}{G(z)^2}+3 G'(z) \Big].
\end{eqnarray}

Additionally by assuming two more conditions for the functions $G$
and $1/G$,
\begin{itemize}
    \item $G$ and $1/G$ are holomorphic in Re$(z) \geq 0$ except at
    $z=0$.
    \item $G$ or $1/G$ are meromorphic at $z=0$.
   \end{itemize}
We can shrink the $C$ contour around infinity, picking up only a
possible contribution from the origin,
\begin{eqnarray}
\label{kine1x5} \langle \Psi Q\Psi \rangle = -\frac{1}{2 \pi^2}
\oint_{C_0} \frac{dz}{2 \pi i} \Big[ -\frac{6 G'(z)}{G(z)}+\frac{3
G'(z)}{G(z)^2}+3 G'(z) \Big],
\end{eqnarray}
where $C_0$ is a contour encircling the origin in the clockwise
direction. The second and third term in the integrand given on the
right hand side of (\ref{kine1x5}) are total derivative terms with
respect to $z$ such that the contour integral of them usually
vanishes. In fact, since we assume the meromorphicity of $G(z)$ at
the origin, these total derivative terms vanish. Now inverting the
direction of the contour $C_0$, we finally obtain
\begin{eqnarray}
\label{kine1x6} \langle \Psi Q\Psi \rangle = -\frac{3}{ \pi^2}
\oint \frac{dz}{2 \pi i} \frac{ G'(z)}{G(z)} .
\end{eqnarray}
Note that to compute this integral (\ref{kine1x6}), we need to
consider a closed contour encircling the origin in the
counterclockwise direction.

\section{The multibrane solution}

It is well known that the spectrum of open string theory contains
tachyons. The presence of these tachyons is a perturbative
consequence of the instability of the D-brane (the space filling
brane) to which open strings are attached \cite{Sen:1999mg}.
According to Sen's conjecture \cite{Sen:1999mh,Sen:1999xm}, there
is a solution of the string field equation of motion, called the
tachyon vacuum solution $\Psi_0$, such that at this vacuum there
is no brane left on which open strings could end. As one
consequence of this statement, the energy computed using the
tachyon vacuum solution $\Psi_0$
\begin{eqnarray}
\label{potencial1} U(\Psi_0) = \frac{1}{2} \langle \Psi_0 Q\Psi_0
\rangle + \frac{1}{3} \langle \Psi_0 \Psi_0 \Psi_0 \rangle
\end{eqnarray}
must cancel the tension of the D-brane
\begin{eqnarray}
\label{tension1} U(\Psi_0)+\mathcal{T}_{D} = 0 ,
\end{eqnarray}
where in some appropriate units ($g_0=1$)\footnote{$g_0$ is the
open string coupling constant. For a more detailed discussion
about these units, we refer to the paper by K. Ohmori
\cite{Ohmori:2001am}.} the tension of the D-brane is given by
\begin{eqnarray}
\label{tension25} \mathcal{T}_{D} = \frac{1}{2 \pi^2} .
\end{eqnarray}
Therefore at the tachyon vacuum solution $\Psi_0$, the energy
(\ref{potencial1}) must have the value
\begin{eqnarray}
\label{tensiontachy1} U(\Psi_0)= - \frac{1}{2 \pi^2}  .
\end{eqnarray}

Suppose that instead of having one D-brane, we have a stack of $n$
coincident D-branes, so that the total energy of this
configuration should be $n \mathcal{T}_{D}$. Now we should ask if
there is a solution $\Psi$ of the string field equation of motion
such that
\begin{eqnarray}
\label{tension1} U(\Psi)+n\mathcal{T}_{D} = 0 .
\end{eqnarray}
Plugging the value of the tension of one D-brane (\ref{tension25})
into this last equation (\ref{tension1}), we obtain
\begin{eqnarray}
\label{potenmultibran1} U(\Psi)= -\frac{n}{2 \pi^2}.
\end{eqnarray}
The solution $\Psi$ which satisfies this condition is known in the
literature as the multibrane solution
\cite{Murata:2011ex,Murata:2011ep}.

Let us derive the energy $U(\Phi)$ in terms of the kinetic energy
evaluated at any general solution $\Phi$. The energy is given by
the sum of the kinetic with the cubic term
\begin{eqnarray}
\label{potencial1sol1} U(\Phi) = \frac{1}{2} \langle \Phi Q\Phi
\rangle + \frac{1}{3} \langle \Phi \Phi \Phi \rangle.
\end{eqnarray}
If we assume the validity of the string field equation of motion
when contracted with the solution itself $\langle \Phi Q\Phi
\rangle +
 \langle \Phi \Phi \Phi \rangle=0$, we can write the energy (\ref{potencial1sol1})
 in terms of the kinetic energy
\begin{eqnarray}
\label{potencial1sol2} U(\Phi) = \frac{1}{6} \langle \Phi Q\Phi
\rangle.
\end{eqnarray}

At this point we are ready to evaluate the energy for a class of
analytic solutions of the form given by equation (\ref{solpsi2}).
Clearly for this solution, using equations (\ref{kine1x6}) and
(\ref{potencial1sol2}), we obtain
\begin{eqnarray}
\label{multipotencilalx1} U(\Psi) = - \frac{1}{2 \pi^2} \oint
\frac{dz}{2 \pi i}  \frac{ G'(z)}{G(z)}.
\end{eqnarray}
A function $G(K)$, which satisfies the holomorphicity conditions
stated in the previous section, is given by
\begin{eqnarray}
\label{GG} G(K)=\Big( \frac{K}{1+K} \Big)^n.
\end{eqnarray}

Since the contour integral (\ref{multipotencilalx1}) is performed
around a closed curve encircling the origin, to compute this
integral we need to write the Laurent series of the integrand
around $z=0$ and pick up the coefficient in front of the term
$1/z$. For a function $G$ defined by equation (\ref{GG}), it turns
out that
\begin{eqnarray}
\label{expandg}\frac{ G'(z)}{G(z)} = \frac{n}{z} + \sum_{m \neq
-1} b_m z^m,
\end{eqnarray}
and consequently the contour integral (\ref{multipotencilalx1})
for this function (\ref{GG}) gives the following result
\begin{eqnarray}
\label{multipotencilalx1yy1} U(\Psi) = - \frac{n}{2 \pi^2},
\end{eqnarray}
which is the expected result for a multibrane solution. Therefore,
in the context of the modified cubic superstring field theory, as
in the case of open bosonic string field theory, we should also
expect a solution which describes the so-called multibrane
configuration \cite{Murata:2011ex,Murata:2011ep}.

\section{Summary and discussion}

We have studied a class of analytic solutions of the string field
equation of motion in the modified cubic superstring field theory.
As in the case of open bosonic string field theory
\cite{Murata:2011ex,Murata:2011ep}, these solutions were
characterized in terms of a single function $F(K)$. We have shown
that this family of solutions can be derived by performing a
suitable gauge transformation over an identity based solution
constructed out of elements in the $KBc\gamma$ subalgebra.

We have analytically evaluated the energy associated to the
solutions characterized by the function $F(K)$. We have shown
that, under certain holomorphicity conditions on the function
$G(K)=1-F^2(K)$, the energy is given in terms of a contour
integral. We have written an explicit form for this function
$G(K)$ and computed the energy associated to this solution. The
result was given by integer multiples of a single D-brane tension.
Therefore, in the context of the modified cubic superstring field
theory, as in the case of open bosonic string field theory, we
should also expect a solution which describes the so-called
multibrane configuration.

Although we have performed analytic computations for evaluating
the energy associated to the class of solutions considered in this
work, it would be nice to confirm our results by employing
numerical techniques such as the curly $\mathcal{L}_0$ level
expansion \cite{Arroyo:2009ec,Arroyo:2011zt} or the usual Virasoro
$L_0$ level expansion scheme
\cite{Moeller:2000xv,Kishimoto:2011zza}. The numerical analysis
should be important, for instance, to check if the solution
behaves as a regular element in the state space constructed out of
Fock states. Specifically the examination of the coefficients
appearing in the $L_0$ level expansion provides one criterion for
the solution being well defined
\cite{Schnabl:2010tb,Takahashi:2007du,AldoArroyo:2011gx}.

Finally, regarding the Berkovits non-polynomial open superstring
field theory \cite{Berkovits:1995ab}, since this theory is based
on Witten's associative star product, its algebraic structure is
mainly similar to both the open bosonic string field theory and
the modified cubic superstring field theory, and hence the
strategy and prescriptions studied in this work should be directly
extended to that theory. However, the construction of analytic
solutions in Berkovits superstring field theory based on elements
in the $KBc\gamma$ subalgebra or extensions of this subalgebra,
remains an unsolved problem.

\section*{Acknowledgements}
I would like to thank Ted Erler, Isao Kishimoto, Masaki Murata and
Martin Schnabl for useful discussions. This work is supported by
FAPESP grant 2010/05990-0.


\end{document}